\documentclass[aps,pre,twocolumn,showpacs,groupedaddress,superscriptaddress,amsmath,amssymb]{revtex4}

\usepackage{graphicx}
\usepackage{amsmath, amssymb}
\usepackage[colorlinks,linkcolor=blue,citecolor=blue]{hyperref}

\newcommand{\second}{\,\text{s}}
\newcommand{\nM}{\,\text{nM}}
\newcommand{\molec}{\,\text{molec}}

\begin{document}

\title{Optimizing periodicity and polymodality in noise-induced genetic oscillators}

\author{Pau Ru\'{e}}
\affiliation{Departament de F\'isica i Enginyeria Nuclear,
Universitat Polit\`ecnica de Catalunya, Edifici GAIA, Rambla de Sant Nebridi s/n, Terrassa 08222, Barcelona, Spain}

\author{G\"urol M. S\"uel}
\affiliation{Green Center for Systems Biology and Department of Pharmacology, University of Texas Southwestern Medical Center, Dallas, TX 75390}

\author{Jordi Garcia-Ojalvo}
\email{jordi.g.ojalvo@upc.edu}
\affiliation{Departament de F\'isica i Enginyeria Nuclear,
Universitat Polit\`ecnica de Catalunya, Edifici GAIA, Rambla de Sant Nebridi s/n, Terrassa 08222, Barcelona, Spain}

\begin{abstract}

Many cellular functions are based on the rhythmic organization of biological processes into self-repeating cascades of events.
Some of these periodic processes, such as the cell cycles of several species, exhibit conspicuous irregularities
in the form of period skippings, which lead to 
polymodal distributions of cycle lengths. 
A recently proposed mechanism that accounts for this quantized behavior is the stabilization of a Hopf-unstable state by molecular noise. 
Here we investigate the effect of varying noise in a model system, namely an excitable activator-repressor genetic circuit, that displays
this noise-induced stabilization effect.
Our results show that an optimal noise level
enhances the regularity (coherence) of the cycles, in a form of coherence resonance. 
Similar noise levels also optimize the multimodal nature of the cycle lengths.
Together, these results illustrate how molecular noise within a minimal gene regulatory
motif confers robust generation of polymodal patterns of periodicity. 
\end{abstract}


\maketitle

\section{Introduction}\label{sec:intro} 

Biological oscillations underlie many physiological functions in cells, from basic processes such as cell growth
and division \cite{Tyson:2001p329} to evolutionary environmental adaptations such as circadian rhythmicity
\cite{Goldbeter:2002p484}. Many circuit architectures have been proposed that explain the observed periodic
behavior in terms of limit-cycle attractors of nonlinear dynamical models \cite{Novak:2008p328}. These limit cycles
exhibit a perfectly periodic behavior, which is only slightly perturbed by realistic levels of biochemical random
fluctuations, or noise, that are unavoidable in cells \cite{Gonze:2002p1214}. In some situations, however, 
cellular oscillations display a degree of variability much larger than what can be obtained from a limit-cycle model
with added noise. This is the case, for instance, of the cell cycle oscillations exhibited by Chinese
hamster cells \cite{Klevecz:1976p829}, fission yeast cells \cite{Sveiczer:1996p662}, and {\em Xenopus laevis}
blastomeres \cite{Masui:1998p581}. In these organisms, cells do not always divide when they are supposed to, 
giving rise to a distribution of cell-cycle periods that is not unimodal, but that exhibits secondary peaks at
multiples of the cell-cycle period.
This quantized behavior cannot be explained by the usual factors responsible for the
heterogeneity of the cell cycle, such as parameter inhomogeneities and the
age-distribution of cells within a population. Those factors, which are undoubtedly present in any
dividing cell population, would only lead to broadening of the period distribution but not to polymodality.
Therefore, detailed mathematical models with a relatively large number of biochemical components
(on the order of 10) have been proposed to explain this behavior
\cite{Sveiczer:2000p853}. In those models, period skipping arises already at the deterministic level
(i.e. in the absence of sources of heterogeneity and inhomogeneity)
\cite{Novak:2001p1331}, while noise is sometimes
considered \cite{Steuer:2004p613} to reproduce the level of irregularity observed in the experiments.
Other striking examples of polymodal cycles embedded in an otherwise oscillatory dynamics were
reported long ago in sensory neurons \cite{Rose:1967p1333} and bacterial motility \cite{Schimz:1992p1334}.

Using the phenomenologies described above as motivation, here we address the general question of
how a limit cycle behavior with polymodal period distribution can arise in minimal oscillator models.
To that end we consider one of the most basic oscillator architectures, namely a two-component
activator-inhibitor
system operating in an excitable regime (close to the oscillatory region) and subject to noise.
We recently showed that such a model system exhibits
noise-induced stabilization of an unstable spiral state
\cite{Turcotte2008}. Due to its excitable character, this model system displays
noise-triggered excursions
away from the stable (rest) state, during which the cell passes through a region near the unstable spiral.
The stabilization mechanism consists in the appearance of oscillations around the unstable state,
due to the stochastic fluctuations. 
As a consequence of these oscillations, the distribution of excursion times away from
the rest state exhibits a marked polymodality:
each noise-induced oscillation around the unstable state introduces a well defined delay
(the oscillation period) in the pulse duration.
Thus noise can explain the polymodality of pulse duration distributions in certain conditions. However,
the (excitable) pulses are triggered by noise to begin with, and thus they are far from occurring
periodically, which would be necessary if this mechanism is to explain the polymodal cell-cycle duration
distributions mentioned above.

Coincidentally, however, systems with excitable dynamics are known to exhibit
enhanced periodicity, or coherence, for an optimal amount of noise: too little noise will elicit pulses only
sparsely, and therefore irregularly, while too high noise will lead to a strong disorder in the dynamics.
A moderate level of noise, on the other hand, is able to evoke pulses frequently, as soon as the refractory
time following the previous pulse (characteristic of all excitable systems) has elapsed, and thus leads to
a substantially periodic behavior, with a period basically given by the refractory time.
Such somewhat counterintuitive effect of noise has been termed
{\em coherence resonance} or, more appropriately, stochastic coherence
\cite{Gang:1993kx,Pikovsky1997,revex2004}.
The goal of this paper is to show that stochastic coherence can be invoked, together with the noise-induced
stabilization effect discussed above, to provide a minimal mechanism for the generation
of polymodal distributions of cycle lengths in an otherwise periodic behavior.
The mechanism requires only a simple genetic activator-repressor motif and
an optimal amount of random fluctuations.
In our setting, the effect of intrinsic molecular noise 
is characterized by using discrete stochastic simulations. 

The level of intrinsic noise is controlled by the cell volume, whose increase (together with
the gene copy numbers) effectively scales up
the numbers of molecular species (thus reducing the noise), while maintaining the concentrations
constant. This approach was recently introduced experimentally in {\em B. subtilis}
\cite{Suel2007}, and has been subsequently used in {\em E. coli} as well \cite{Fischer-Friedrich:2010vn}.
From the theoretical side, system-size effects have been seen to lead to stochastic-resonance
\cite{Pikovsky:2002ly,al:2001zr} and stochastic-coherence \cite{al:2003ve} effect through their control
of the effective noise intensity perceived by the system.

Our results show that noise, besides
enhancing the regularity of the pulse activations, also optimizes polymodality
in the system's response. Furthermore, optimization of periodicity and polymodality
are achieved when noise levels are comparable. Thus, when the coherence
of the excitable pulses is maximized, so is the probability that the pulses undergo
oscillations around the unstable spiral state. There is a range of noise levels for
which optimization holds. Together, these results show that noisy
activator-repressor genetic circuits can naturally behave as polymodal oscillators.

%%%%%%%%%%%%%%%%%%%%%
\section{Model}\label{sec:model}

\begin{figure}[htbp]
\centering
\includegraphics{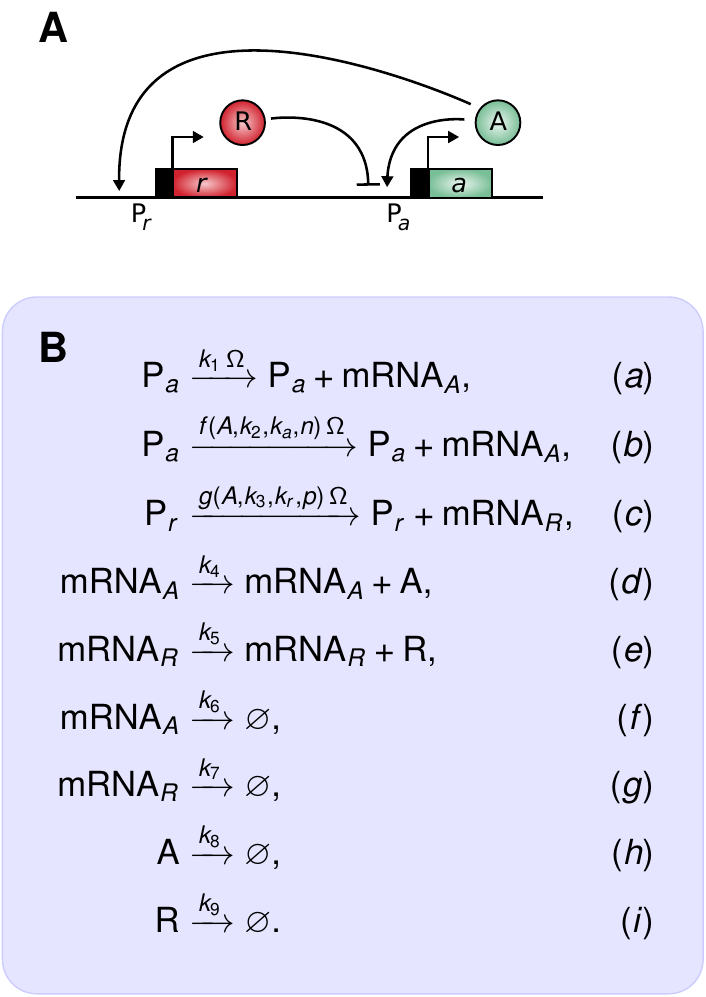}
\caption{\label{fig:circuit}
(A) The genetic motif investigated here consists of two genes, \emph{a} and \emph{r}, coding for activator, $\text{A}$, and a repressor, $\text{R}$, proteins respectively. 
The activator protein binds to the promoters $\text{P}_a$ and $\text{P}_r$ of the genes \emph{a} and \emph{r} respectively, and the repressor protein binds to  $\text{P}_a$ competitively inhibiting its activation by $\text{A}$.
(B) Set of reactions describing the genetic circuit.
Reaction (a) corresponds to the leaky/constitutive transcription of activator mRNA. Regulated transcription of the activator and repressor mRNA species is encoded in reactions (b) and (c). Protein translation for the activator and repressor is represented by reactions (d) and (e), respectively,
and degradation reactions for the mRNA and protein species are (f), (g), (h) and (i).
}
\end{figure}

We now describe our model system using the terminology of gene regulation circuits,
although the results obtained are applicable to any activator-inhibitor system.
The genetic circuit (see Figure~\ref{fig:circuit}) is a simple
two-component system where an activator protein, $\text{A}$, binds to and activates its own promoter, $\text{P}_a$, and the promoter of a repressor species $\text{P}_r$. 
The repressor component, $\text{R}$, in turn,  inhibits the expression of the activator species by competitively binding to the promoter $\text{P}_a$.
This system is a prototypical transcriptional activator-repressor genetic circuit where the activator species forms a 
direct positive feedback loop through its auto-regulation and an indirect negative feedback loop by means of the activation of its own repressor \cite{Tsai:2008p907,Stricker:2008p373,Tigges:2009p398}. 
The dynamics of this system is highly nonlinear, due to the cooperative nature of the regulated
transcription processes. 
Specifically, the kinetics of the regulated activation of the $\text{P}_a$ promoter is described by a Hill function
with cooperativity $n$:
\begin{equation*}
f(A, R, k_2, K_a, K_i, n, m) = \frac{k_2A^n}{A^n+ K_\text{eff}^n},\\
\end{equation*}
where $A$ represents the activator concentration in the cell, and the effective activation
threshold $K_{\text{eff}}$ 
depends on the concentration $R$ of repressor through the expression
${K_{\text{eff}}}^n = {K_a}^n\left(1  + \left(\frac{R}{K_i}\right)^{m}\right)$.
In this equation, the term $K_i$ accounts for the competitive inhibition exerted by $\text{R}$, the net effect of which is to increase the promoter's activation threshold.
The kinetics of regulated transcription of the repressor mRNA is described by a simple activation Hill equation
\begin{equation*}
 g(A, k_3, K_r, p) = \frac{k_3}{1 + \left(\frac{K_r}{A}\right)^p }.
\end{equation*}
with constant activation threshold $K_r$.

The values of the reaction rates used in the simulation shown below are given in Table~\ref{tab:params}.
The values of the parameters are within reasonable biological ranges for a gene regulation circuit.
In particular, the values chosen for the transcrition, translation, and degradation rates, and for the
activation and inhibition threshold concentrations, are of the same order of magnitudes of previous
studies that involved qualitative comparison and careful validation with experimental measurements
\cite{Suel2006,Suel2007,Cagatay2009}.
Note that the rates of the transcription reactions in Fig.~\ref{fig:circuit} are proportional to a factor $\Omega$.
This parameter is a global scaling factor that depends on the size of the cell.
Specifically, $\Omega=V\cdot N_A$, where $V$ is the cell volume and $N_A$ is Avogadro's number.
In that way, $\Omega$ relates the species concentrations with the number of molecules:
if $n_\text{A}$ is the number of molecules of $\text{A}$ and $A$ its concentration, then $n_\text{A} = \Omega \, A$.
We will also consider that the transcription strengths are proportional to $\Omega$, which
is a generalization to the continuous of the assumption that
the gene copy number increases with cell size. This happens when a cell is prevented from dividing
but not from replicating its DNA, as happens in certain bacterial mutants \cite{Suel2007}. Under
those assumptions, the level of
molecular noise increases monotonically with the inverse of the system size, $\Omega^{-1}$
\cite{Gillespie1977}. In the following, we will use the parameter $\Omega^{-1}$ to characterize the
levels of noise in the system.

\begin{table}
\begin{center}
\begin{tabular}{||c|c||c|c||}
\hline
$k_1$ &  0.00625\nM/(\second$\cdot$\molec) & $k_6$ & 0.05\second$^{-1}$\\
$k_2$ & 0.5\nM/(\second$\cdot$\molec) &$k_7$ &  0.05\second$^{-1}$\\
$k_3$ & 0.05\nM/(\second$\cdot$\molec) &$k_8$ & 0.001\second$^{-1}$\\
$k_4$ & 2\second$^{-1}$ &$k_9$ & 0.0001\second$^{-1}$ \\
$k_5$ & 2\second$^{-1}$ &$K_a$ & 5000\nM\\
$K_r$ & 9000\nM  & $K_i$ & 5000\nM \\
$n$ & 2 & $m$&2    \\
$p$ & 4	& &\\
\hline
\end{tabular}
\caption{\label{tab:params} Values of the reaction rates used in the stochastic simulations
of the circuit represented in Fig.~\protect\ref{fig:circuit}.}
\end{center}
\end{table}
A continuous model of the circuit for the case of negligible fluctuations can be derived for the set of reactions
listed in Fig.~\ref{fig:circuit}B
\cite{Alon:2006ys}.
Let $a$, $r$ denote the concentration of activator and repressor mRNA molecules and $A$, $R$ the protein concentrations. Applying standard kinetics rules to the reactions listed in Fig.~\ref{fig:circuit}B leads
in a straightforward way to the following coupled ordinary differential equations:
\begin{equation}\label{eq:reaction-rates}
\begin{split}
\frac{da}{dt} &=k_1 + k_2 \frac{A^n}{A^n + {K_a}^n + \left({\gamma_r} R\right)^m} - k_6 a, \\
\frac{dr}{dt} &=k_3 \frac{A^m}{{K_r}^p + A^p} - k_7 r,\\
\frac{dA}{dt} &= k_4 a - k_8 A,\\
\frac{dR}{dt} &= k_5 r -k_9 R.
\end{split}
\end{equation}
where ${\gamma_r}^m = \frac{{K_a}^{n}}{{K_i}^m}$.
Note that this system of equations is independent of the system size.

Further assuming a separation of mRNA and protein time-scales, the former ones (being much smaller) can be adiabatically eliminated, and the system can be reduced to two coupled ordinary differential equations
\begin{equation}\label{eq:rre2d} 
\begin{split}
\frac{dA}{dt} &=\alpha + \beta_1 \frac{A^n}{A^n + {K_a}^n + \left(\gamma_r R\right)^m} - \lambda_1 A \\
\frac{dR}{dt} &=\beta_2 \frac{A^p}{{K_r}^p + A^p} - \lambda_2 R
\end{split}
\end{equation}
where
\begin{equation}\label{eq:param_conversion}
\begin{alignedat}{3}
\alpha &= k_1 k_4/k_6,\qquad \qquad       & \\
\beta_1 & = k_2 k_4/k_6, 		 & \beta_2&= k_3 k_5/k_7,\\
\lambda_1&=k_8,			& \lambda_2&=k_9.
\end{alignedat}
\end{equation}
While in this work the model dynamics is obtained by exact discrete stochastic simulation of the chemical kinetics
\cite{Gillespie1977} described in Figure~\ref{fig:circuit}B, its qualitative dynamical aspects are described
with the planar system of differential equations (\ref{eq:rre2d}).
We note that, as in the case of Eqs.~(\ref{eq:reaction-rates}), our final deterministic model
given by Eqs.~(\ref{eq:rre2d}) does not depend on the system size, and thus the deterministic
dynamics will be unchanged as noise levels vary. This is due to the fact that the system size factor
$\Omega$ rescales the levels of all proteins (and thus the absolute activation and inhibition thresholds)
in the same manner as the transcription rates, while keeping the concentrations unchanged.
In that way, the average concentration dynamics of the model does not depend on noise, but only the
variances of the concentrations do.

%%%%%%%%%%%%%%%%%%%%%%%%%%
\section{Results}\label{sec:results}
\subsection{Deterministic excitable dynamics and effects of molecular noise}

\begin{figure}[htbp]
\includegraphics{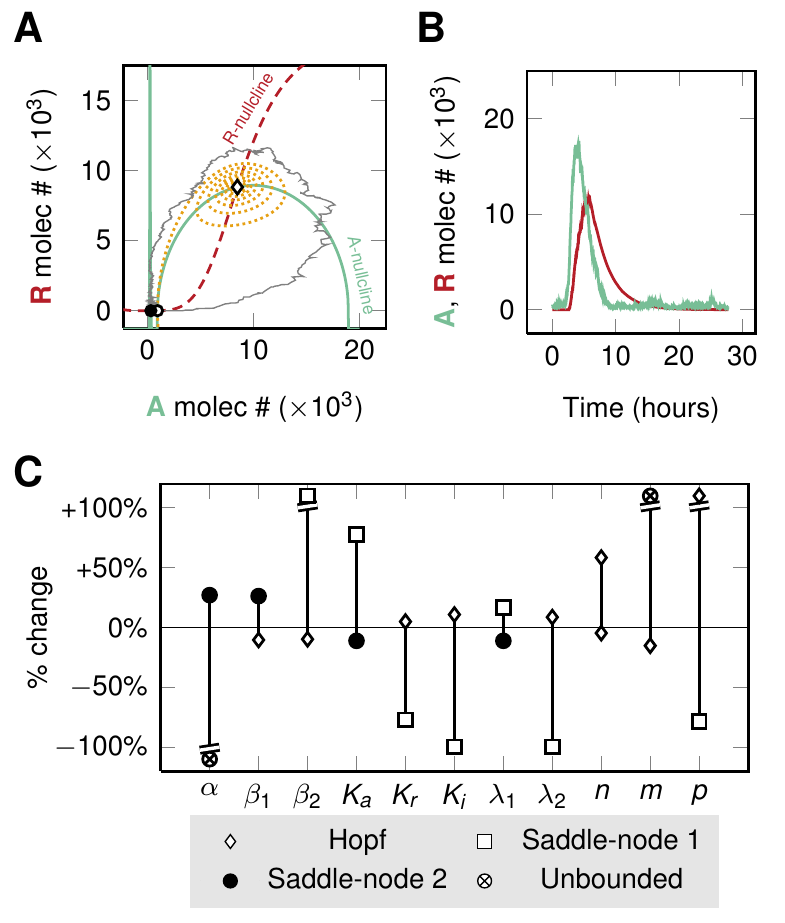}
\caption{\label{fig:dynamics} %
(Color online) (A) Phase portrait representing the dynamics of the system of equations \eqref{eq:rre2d}.
The nullclines for $A$ and $R$ are represented by solid (green online) and dashed (red online) lines.
The system has three equilibrium points: a stable node (black circle), a saddle point (white circle) and 
an unstable focus (white diamond). The stable manifold of the saddle point (dotted line, yellow online)
introduces a threshold of excitability. A trajectory obtained by discrete stochastic simulations
($\Omega=1\molec/\nM$) is shown in black. The random fluctuations of this system make it possible to cross the separatrix 
and initiate a large excursion in the phase plane.
(B) The same trajectory is plotted as a time course of the protein species. When the threshold of excitability
is crossed, a transient pulse in the number of activator and repressor molecules is produced.
(C) Sensitivity of the model to single parameter changes. Vertical lines show the range of parameter values for which excitability is maintained (in $\%$ change from the values in Table~\ref{tab:params}). The symbols at the end of the vertical
lines indicate the type of bifurcation leading to loss of excitability, or whether the range is unbounded (see legend and main text). Broken bars indicate much larger ranges than the one indicated by the vertical axis.
}
\end{figure}

The interplay between the auto-activation positive feedback loop of $A$ 
on itself and the negative feedback loop formed by the activator and the repressor
allow for a wide range of rich dynamics. 
In particular, for the set of parameters given in Table~\ref{tab:params} and Eq.~\eqref{eq:param_conversion}
the system is excitable.
The phase portrait depicting the nullclines of this system for those parameter values is shown
in Fig.~\ref{fig:dynamics}A. 
The system has three equilibrium points: a stable node, a saddle point and an unstable focus. 
In the absence of noise, the system rests in the only stable state, which in this case corresponds to low numbers
of both activator and repressor molecules.
Small perturbations from this stable point vanish exponentially and the system quickly recovers the rest state. 
However, the stable manifold of the saddle point (the separatrix, dotted spiral line in Fig.~\ref{fig:dynamics}A)
is an excitability threshold beyond which perturbations evoke a large excursion through phase space, passing around the unstable focus and back to the stable point avoiding to cross the separatrix~\cite{Izhikevich2006}.  
The occurrence of this excitability cycle can be understood as follows: a sufficiently large initial amount of activator molecules
triggers the positive feedback loop and leads to a large pulse of activator molecules. The increasing levels of activator switch on the production of repressor proteins (repressor pulse) which, in turn, shut down the production of activator. Finally, the amounts of activator and repressor decay due to linear degradation/dilution.
This transient response is characterized by a refractory time, which is the duration of the cycle from the triggering event to the recovery of the rest state.

Stochastic fluctuations due to intrinsic noise can destabilize the rest state by randomly crossing the excitability threshold and hence generating pulses of activator and repressor protein levels. Figure~\ref{fig:dynamics}A shows an excursion in phase space triggered by stochastic fluctuations, and Fig.~\ref{fig:dynamics}B shows the corresponding time course for both the activator and the repressor. 

The excitable regime in which this system operates arises when the system is close to a bifurcation point 
beyond which the dynamics has the form of a limit cycle~\cite{Rue2010}. In addition, the noise induced stabilization 
of the unstable state emerges close to a Hopf bifurcation~\cite{Turcotte2008} beyond which the system becomes bistable.
Despite these constraints in the parameter values, the activator-represor system presented here is robust to parameter 
changes, as shown in Figure~\ref{fig:dynamics}C.
This Figure depicts the main bifurcations from excitability that the system undergoes as the kinetic parameters are 
varied one by one. 
In particular, the codimension-one bifurcations found are i) the stabilization of the unstable spiral via a Hopf 
bifurcation; ii) the collision of the saddle and the node defining the resting state (Saddle-node 1 in the 
Figure~\ref{fig:dynamics}C); and iii) the collision of the saddle and the unstable state (Saddle-node 2). 
This analysis shows that the more sensitive parameters are the activator degradation rate, $\lambda_1$, for which 
the excitability is maintained for a global range of variation of $27.8\%$ and the the maximum regulated transcription 
rate, $\beta_1$, for which a global variation of $36.7\%$ is possible without losing excitability.

\subsection{Polymodality in the cycle duration depends on the level of intrinsic noise}
% burst duration

\begin{figure}[htbp]
\includegraphics{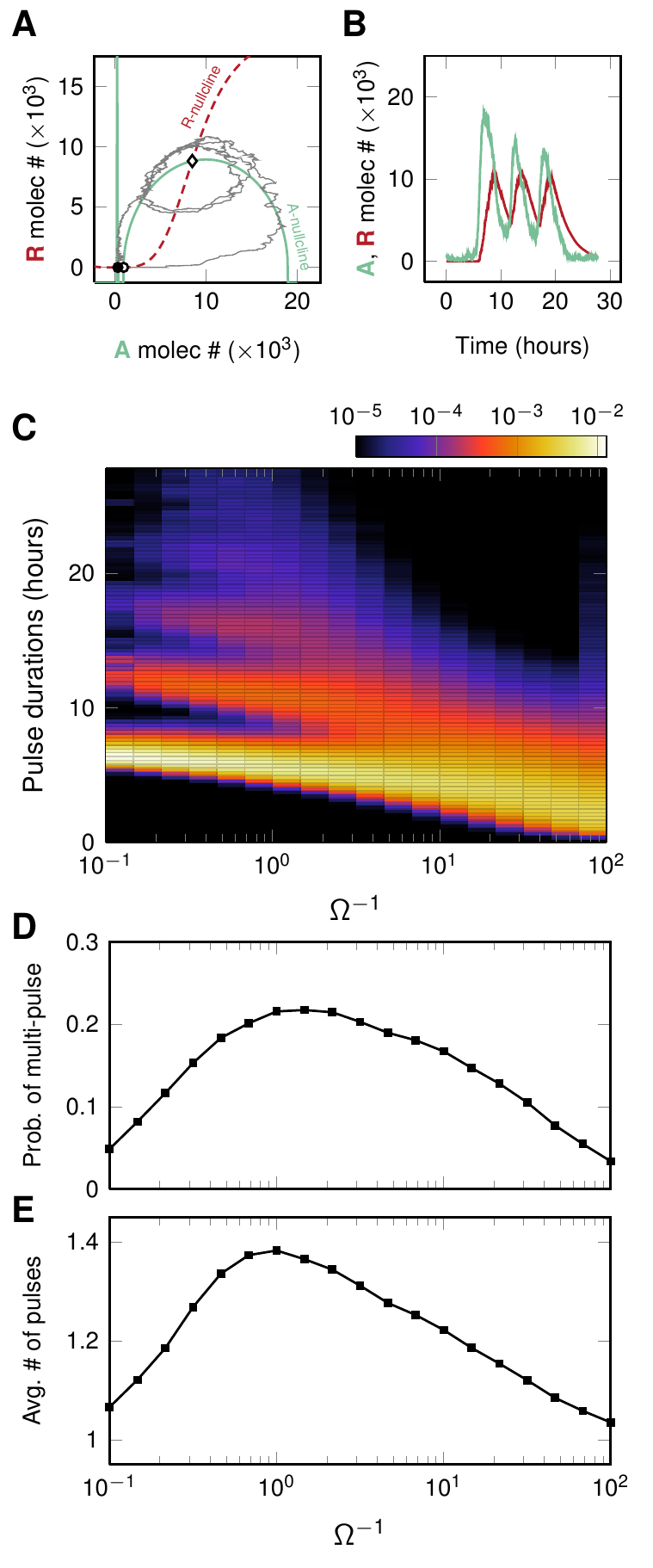}
\caption{\label{fig:nis}
Molecular noise induces bursts with multiple pulses. 
(A) A burst trajectory obtained with discrete stochastic simulations ($\Omega=1\molec/\nM$) is depicted in the phase plane. 
Fluctuations due to molecular noise synergistically interact with the dynamics around the focus and can temporarily trap the system in an area around the otherwise unstable state.
(B) Time course for the number of activator and repressor proteins for the burst shown in (A).
The stabilization of the active state is characterized by the oscillations in the numbers of molecules.
(C) Histograms of the burst durations for varying levels of the intrinsic noise ($\Omega^{-1}$). Color is coded in logarithmic scale to emphasize the existence of polymodality.
(D) Probability of generating a cycle with multiple pulses plotted against the noise level. For intermediate values of the
noise strength the system reaches a maximum probability of generating bursts with more than one pulse. 
(E) Average number of pulses per burst of activity as a function of $\Omega^{-1}$.}
\end{figure}

As already mentioned, the sporadic generation of pulses of activity is not the only effect caused
by molecular noise in this system. As shown in Figs.~\ref{fig:nis}A and \ref{fig:nis}B, noise also
stabilizes the unstable state and generates bursts, or cycles of multiple pulses.
This is due to the fact that noise causes the trajectories traveling around the unstable spiral point
to cross over the stable manifold of the saddle (dotted line in Fig.~\ref{fig:dynamics}A). This leads
to the trajectory getting trapped orbiting around the unstable fixed point for an integer number of
cycles, thus generating a polymodal distribution of pulse durations \cite{Turcotte2008}.

In order to characterize this noise-induced polymodality, we have computed the cycle durations for varying 
levels of molecular noise (which increase as the system size $\Omega$ decreases). 
Figure~\ref{fig:nis}C shows the normalized histogram of burst durations for different values of $\Omega^{-1}$. 
For large system sizes (small $\Omega^{-1}$, small molecular fluctuations), the histogram shows multiple modes with clearly defined peaks.
Each of the modes of the histogram obviously corresponds to a class of burst with a definite number of pulses.
The single pulse cycle (corresponding to the first mode in the histogram) is, by far, the most probable case.
For intermediate noise levels, the secondary peaks, corresponding to cycles with multiple pulses, get both higher
and wider. 
If noise is increased further, the polymodal character of cycle durations is lost and just a single wide peak remains.
%on one side, the polymodality is sharp but the first mode is much higher than the rest (the probability of getting multi-pulse burst is very low) and, on the other, noise kicks in and overpasses the spiral dynamics around the unstable focus.

% pulses per burst

These results insinuate a new resonance-like effect in which growing levels of noise increase the polymodality in the pulse-duration distribution, until it reaches a maximum and starts declining again.
In order to assess this resonant effect, we have computed the probability of having more than one pulse in one
burst (Fig.~\ref{fig:nis}D), together with the
number of pulses per activation burst (Fig.~\ref{fig:nis}E) as a function of increasing noise levels. 
Here, the pulses per cycle are computed by counting the number of complete cycles around the unstable focus. 
This method was found to be a very robust way to compute the number of pulses in the presence of large random fluctuations. 
Both plots clearly show the predicted resonance effect, with the optimal degree of polymodality arising at
a value of $\Omega\sim1\molec/\nM$. 

The reasoning behind this resonant effect can be stated as follows.
For large system sizes (small noise), noise is large enough to trigger excitable pulses, but it is too small
to easily induce crossings of the trajectory beyond the stable manifold of the saddle. Thus the fraction of bursts in
which there is more than one cycle is small (and the average number of cycles is close to one).
On the opposite side, for small system sizes, the random fluctuations are too large to maintain the coherence of
the oscillations around the unstable spiral, and the burst duration is no longer quantized, but is widely variable
and with small mean, since it is easy for the trajectory to escape the area near the unstable spiral and relax
back to the neighborhood of the rest state.  
For intermediate system sizes, on the other hand, noise is large enough to induce frequent crossings of the
spiral's stable manifold, but small enough to maintain the coherence of the noise-induced oscillations, and
thus polymodality is maximized.

\subsection{Noise modulates the regularity/coherence of the oscillations}
\begin{figure}[htbp]
\includegraphics{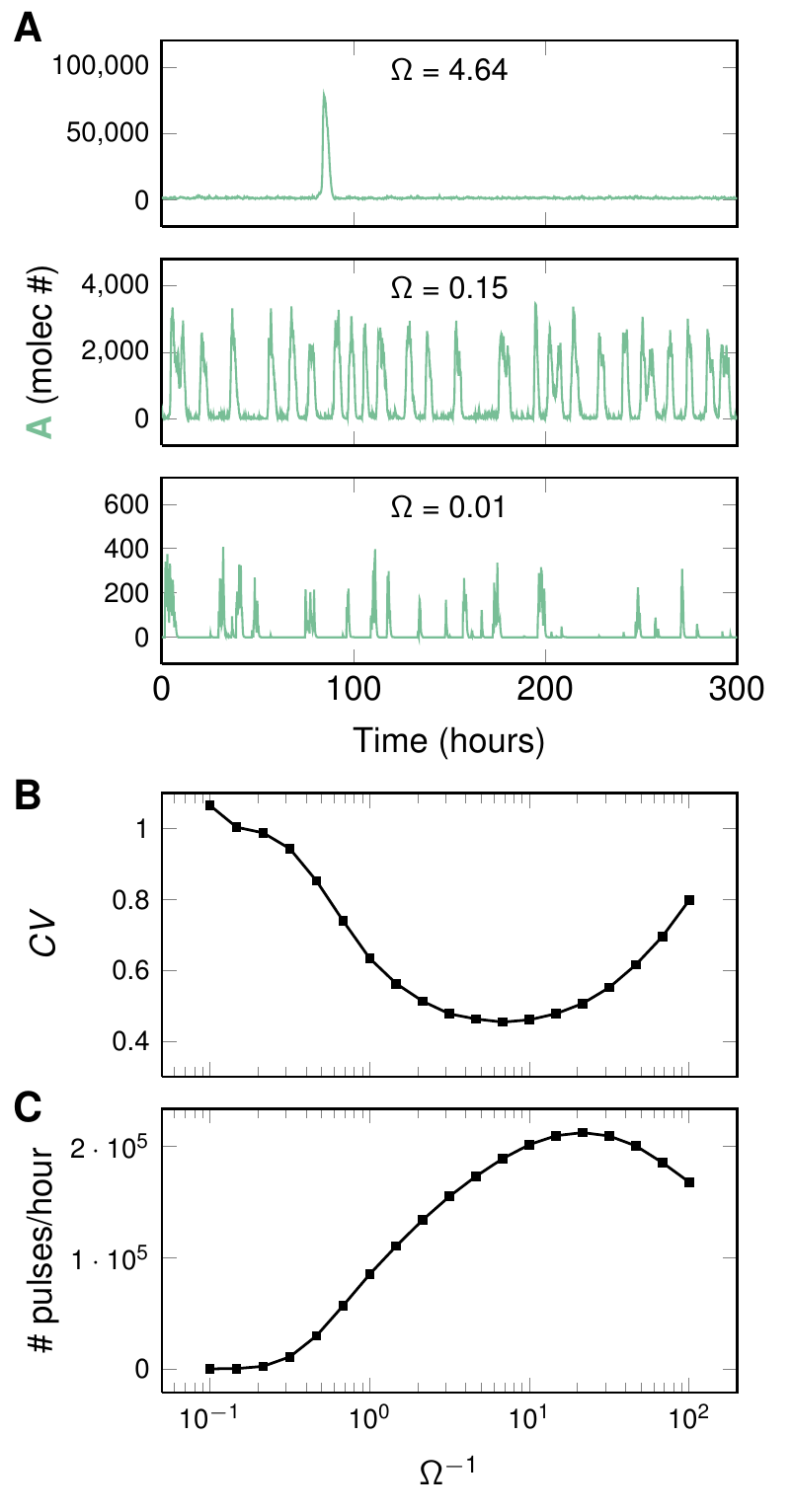}
\caption{\label{fig:cr}
(A) Temporal evolution of number of activator molecules for three different values of the system size. Note
how the peak number of molecules decreases as the system size decreases.
(B) Coefficient of variation of the intervals between consecutive cycle initiations, plotted versus $\Omega^{-1}$.
(C) The average rate of cycle initiation as a function of the noise level. For large system sizes, the fluctuations are not large enough to initiate the cycles. For small system sizes the decay in the pulsing rate is mainly due to the discrete nature of the chemical reactions and the small amount of mRNA species.
}
\end{figure}

%We have 
Let us now turn our attention to the ability of the system to generate regular cycles. 
Not being a genuine genetic oscillator but an excitable system, pulses in this circuit
are in principle randomly generated by noise. 
In this scenario, we want to establish whether the level of noise has an impact in the regularity of pulse initiation.
This effect is already made evident by visual inspection of the time traces of the activator species
for different values of $\Omega$, as shown in Figure~\ref{fig:cr}A.
The three panels in this Figure display time traces of the the activator species for three different levels of molecular noise (increasing noise from top to bottom). The dynamics of the system in this three panels are qualitatively different. While the top and bottom panels show bursts of activity at very irregular time intervals, the middle panel presents a quite regular pattern of cycles.
This plot already hints at a second noise-dependent resonance-like effect, according to which
the regularity of excitable pulses is maximized for an intermediate amount of noise, what is
known as coherence resonance or stochastic coherence \cite{Pikovsky1997,revex2004}. 
In order to quantify this effect, we compute the coefficient of variation ($CV$) of the time between bursts of protein concentration, a reliable regularity measure.
The CV is defined as the standard deviation normalized to the mean
$$CV=\frac{\sqrt{\langle T^2\rangle - {\langle T\rangle}^2}}{\langle T \rangle},$$
where the random variable $T$ denotes the duration between bursts of activity.
This measure is routinely used as a quantifier of noise-induced regularity in coherence-resonance
studies \cite{revex2004}.
In a perfectly periodic regime where the pulses were equally spaced in time, the coefficient of variation would be exactly zero. Conversely, in a completely random regime with pulses following poissonian statistics (exponential waiting times), the coefficient would take the value of one (the standard deviation being equal to the mean) or even higher (hyper-exponential waiting times). Thus, the smaller the value of CV, the closer
the system is to operate in a coherent regime. Other quantifiers of coherence resonance, such as
the correlation time of the dynamics, can be used, but lead to the same conclusions
\cite{Pikovsky1997}.
Here, the CV is estimated from the simulated time courses using a robust thresholding method.

The coherence resonance effect is revealed in Fig.~\ref{fig:cr}B, where the $CV$ is plotted against the system size. 
For small amounts of intrinsic noise the system mainly remains in the basal stable state, with some sporadic pulses appearing randomly in time (see also Figure~\ref{fig:cr}C, which shows that the pulsing rate approaches 0 for small noise).
This results in a $CV$ value around 1.
As the system size is decreased, the effect of intrinsic noise increases and the system pulses at a higher pace. Here is where the refractory time enters the game, as it poses a limit in the pulsation frequency (the system cannot undergo a new cycle if it has already started one). Thus, temporal regularity appears as a synergistic effect involving the dynamics of the system and the intrinsic noise. In particular, the maximum regularity in the oscillations (a minimum CV) appears at $\Omega=0.147\molec/\nM$.
Further reducing the system size causes a reduction in the regularity of the oscillations. 
This loss of coherence is due to two main causes: first, large amounts of noise destroy the excursion paths, thus generating a variety of incomplete pulses and eliminating the system's eigenfrequency dictated by the refractory time
\cite{revex2004}.
The second cause of coherence loss, which is not common in standard coherence resonance, is
the appearance of periods of silencing where the activity of the system is completely shut down (see bottom panel in Figure~\ref{fig:cr}A). These periods of silencing are due to the fact that when the cell size is very small, the
number of molecular species is very small. In particular, the number of mRNA molecules falls frequently
to zero,
resulting in the total absence of protein expression during relatively large time intervals. Such low levels
of mRNA are not uncommon in cells, as has been recently shown experimentally in {\em E. coli}
\cite{Taniguchi:2010p1097}.

\subsection{Polymodality affects regularity}

We have shown that this circuit displays two apparently opposing effects caused by intrinsic noise. 
On one side, noise increases the variability in the duration of the cycles in a quantized manner.
On the other side, it reduces the variability in the cycle initiation times. Thus, it is reasonable to think that these effects might affect one another.
Here we show how polymodality in the duration of the cycle poses a limit in the temporal coherence attained by the genetic oscillator.
For this purpose, we first assume that, for a given range of noise we reach a perfect timing of the cycles, where each cycle follows the next without delay.
In this hypothetical case we can estimate the loss of temporal coherence (in the $CV$ sense) attributable to the cycle duration polymodality by taking into account the probabilities of obtaining a cycle with a particular number of bursts. 
In this case, in which a new cycle starts just after the previous one, the average time between cycles is $\langle T_\text{pm} \rangle = \sum_{i\geq 1} p_i\, T_i$, 
where $p_i$ is the probability of getting a burst with exactly $i$ pulses ($i>0$) and $T_i$ is the average duration of a burst with $i$ pulses.
We can also estimate the variance by further assuming that all the cycles with $i$ pulses have a length of exactly $T_i$ (zero variance among them): ${\sigma_\text{pm}}^{2}=\langle {T_\text{pm}}^2 \rangle-\langle T_\text{pm}\rangle^2 = \sum_{i\geq 1} \left(p_i\, {T_i}^2\right)-\langle T_\text{pm}\rangle^2$. This allows us to calculate the contribution of the polymodality,
$CV_\text{pm}$, to the total coefficient of variation of the time between initiations.

\begin{figure}[htb]
\includegraphics{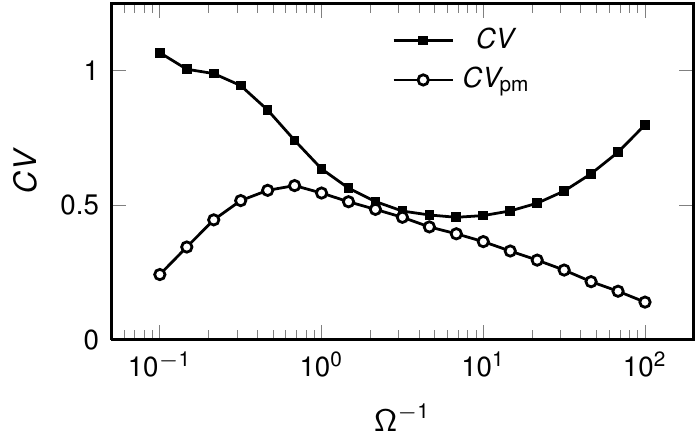}
\caption{\label{fig:cv_pm} 
Coefficient of variation of the inter-burst times due to polymodality, $CV_\text{pm}$ (white circles), compared to the
total $CV$ of the dynamics (black squares, from Fig.~\ref{fig:cr}B). For intermediate values of noise the $CV$ of
the oscillator becomes close to $CV_\text{pm}$.} 
\end{figure}

Figure~\ref{fig:cv_pm} compares $CV_\text{pm}$ (white circles) with the total 
 $CV$ (black squares, see also Fig.~\ref{fig:cr}B) for varying noise levels. 
In this Figure, $CV_\text{pm}$ has been computed using values for the probabilities $p_i$
that were estimated from the time traces of the stochastic simulations (see Fig.~\ref{fig:nis}D
and accompanying text). 
In addition, the values for $T_i$ have been fitted to the formula $T_i/T_1= 1+ (i-1)\delta $
(resulting in $\delta=0.75$).
A comparison between the white circles in Fig.~\ref{fig:cv_pm} and the result of Fig.~\ref{fig:nis}E shows
that, for continuous pulsing (noise generates a pulse as soon as the
refractory time from the previous pulse is over), the regularity of the pulses drops ($CV_\text{pm}$ increases),
as the polymodality is enhanced. Thus, in the regime of continuous pulsing there is an evident tradeoff
between cycle length polymodality and temporal coherence.
Finally, Fig.~\ref{fig:cv_pm} also shows that around the minimum $CV$ the system is close to the regime of
constant cycling. Thus in that case basically all the remaining irregularity (note that $CV$ does not decay to zero)
is due to the polymodality. Therefore, polymodal behavior establishes an upper bound for the regularity
of the system's dynamics.

\subsection{Noise induced polymodality and regularity robustly coexist}

\begin{figure}[htb]
\includegraphics{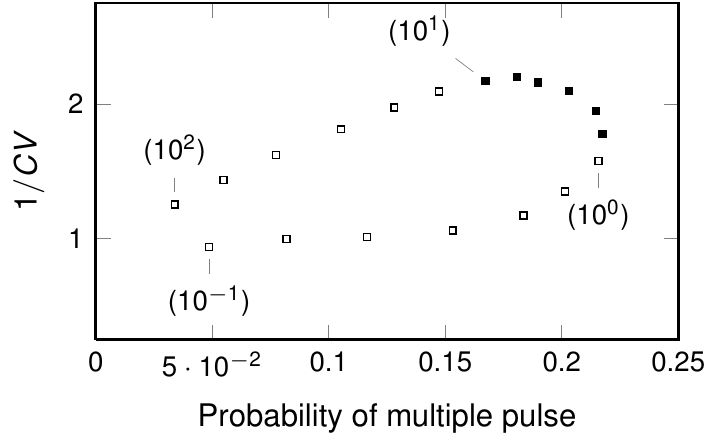}
\caption{\label{fig:optimal_noise}  A range of noise levels optimize regularity and polymodality almost
simultaneously. Regularity ($1/CV$, from Fig.~\ref{fig:cr}) is plotted against polymodality
(probability of multiple pulses in a burst, from Fig.~\ref{fig:nis}D) for varying levels of molecular noise.
Labels indicate the inverse of the system sizes ($\Omega^{-1}$) for some of the points. Black squares
correspond to noise levels for which the system is both largely polymodal (probability $\geq 0.15$)
and regular ($CV<0.6$).}
\end{figure}

An important question is how the noise levels that optimize coherence and polymodality compare to one
another. A comparison between Figs.~\ref{fig:nis}D and \ref{fig:cr}B reveals that, for the parameter values
chosen, there is one order of magnitude difference between the two optimal noise levels. However, we
can still say that the two effects displayed by this simple genetic model take place for overlapping ranges
of molecular noise. This can be argued from Fig.~\ref{fig:optimal_noise}, which plots together the measures
of polymodality (in the x-axis) and regularity (in the y-axis) for varying amounts of noise (milestones labeled
in some data points in the figure).
The figure shows that as noise increases, both the regularity and polymodality increase, and
there is a range of noise
levels spanning over an order of magnitude (black squares in the figure), for which both magnitudes are equally high,
before decreasing again as noise is further increased. Thus, one can say that an optimal level of noise maximizes
almost simultaneously, and for a relatively wide range of noise amplitudes, both the regularity and the polymodality
of the dynamics of the activator-repressor module.

%%%%%%%%%%%%%%%%%%%%%%%%%%
\section{Discussion}\label{sec:discussion}

Cellular processes regulated by genetic components are subject to large amounts of random fluctuations. 
In the face of this fact, it is appealing to conjecture that, rather than simply trying to filter out noise, certain
cellular mechanisms have evolved to cope with random fluctuations, and in some cases even rely on
them for function. 
In the last decades many noise-induced phenomena in physical systems have been described, both
theoretically and experimentally.
Strikingly, noise can in some cases increase order in the dynamics~\cite{Sagues:2007uq} and
play a constructive role in nonlinear systems.
It is also becoming evident in recent years that molecular noise has an impact on the dynamics underlying many
biological processes \cite{Cagatay2009,McKane:2005p1335,Alonso:2007p1337,EldarElowitz2010,Kittisopikul:2010fk}.

Here we have shown that intrinsic noise is able to turn a simple activator-repressor genetic circuit into
an oscillator with non-trivial statistical properties, reflected in a polymodal distribution of cycle
durations embedded in a relatively strongly periodic sequence.
A similar effect has been reported in coupled excitable elements
\cite{Volkov:2003p1332,Koseska:2007p899}. Here, in contrast, we show
that the phenomenon can arise in single excitable elements. The role of noise in our system is two-fold.
On the one hand, it stabilizes an unstable spiral point by inducing oscillations around it, which
increases the duration of phase-space excursions in a quantized manner. Furthermore, the resulting
polymodal character of the dynamics is enhanced for an intermediate noise level. The second role of noise
is to enhance coherence in the pulse initiation times, which occurs via a standard coherence
resonance effect, characteristic of excitable systems subject to noise \cite{revex2004}. This
double optimization provides a relatively simple mechanism for the emergence of 
polymodal behavior in genetic oscillators.

\section*{Acknowledgements}

We thank Ekkehard Ullner for useful discussions. This work has been financially supported by
the Ministerio de Ciencia e Innovaci\'{o}n (Spain, project FIS2009-13360).
P.R. is supported by the FI programme from the Generalitat de Catalunya.
G.M.S. acknowledges support by grants NIH NIGMS RO1 GM088428, Welch Foundation (I-1674) 
and James S. McDonnell Foundation (220020141). G.M.S. is a W. W. Caruth, Jr. Scholar of 
Biomedical Research.
J.G.O. also acknowledges financial support from the ICREA foundation.

\bibliographystyle{apsrev}

\end{document}